\documentclass[12pt,preprint]{aastex}
\usepackage{natbib}
\begin{document}
\newcommand{\be}{\begin{equation}}
\newcommand{\bel}[1]{\begin{equation}\label{eq:#1}}
\newcommand{\ee}{\end{equation}}
\newcommand{\bd}{\begin{displaymath}} %like \be, but doesn't put in eqn. number
\newcommand{\ed}{\end{displaymath}}   %like \ee, but doesn't put in eqn. number
\newcommand{\bea}{\begin{eqnarray}}
\newcommand{\beal}[1]{\begin{eqnarray}\label{eq:#1}}
\newcommand{\eea}{\end{eqnarray}}
\newcommand{\e}[1]{\label{eq:#1}}
\newcommand{\eqref}[1]{\ref{eq:#1}}

\newcommand{\bfr}{{\bf r}}
\newcommand{\bfrp}{{\bf r'}}
\newcommand{\Scal}{{\cal S}}
\newcommand{\Jcal}{{\cal J}}
\newcommand{\Bcal}{{\cal B}}
\newcommand{\Jbar}{{\bar J}}
\newcommand{\Sbar}{{\bar S}}
\newcommand{\Jcalbar}{{\bar \Jcal}}
\newcommand{\Ibg}{{I_{\rm bg}}}
\newcommand{\etal}{et~al.}

\slugcomment{ }

\shorttitle{HypercompacHII regions}
\shortauthors{Eric Keto}

\title{The formation of massive stars: accretion, disks, and the
development of hypercompact HII regions}

\author{Eric Keto }
\affil{Harvard-Smithsonian Center for Astrophysics, 60 Garden St., Cambridge, MA 02138}

\begin{abstract}
The hypothesis that massive stars form by accretion can be investigated
by  simple analytical calculations that describe the effect that the formation of
a massive star has on its own accretion flow.
Within a simple accretion model that includes angular momentum,
that of gas flow on ballistic trajectories around a star,
the increasing ionization of 
a massive star growing by accretion produces a three-stage evolutionary sequence.
The ionization first forms
a small  quasi-spherical
HII region gravitationally trapped within the accretion flow. At this stage the flow of
ionized gas is
entirely inward. As the ionization increases, 
the HII region transitions to a bipolar 
morphology in which the inflow is replaced by outflow within a narrow
range of angle with
about the bipolar axis.
At higher rates of ionization,
the opening angle of the outflow region progressively increases.
Eventually, in the third stage, the accretion is confined
to a thin region about an equatorial disk.
Throughout this early evolution, the HII region is of hypercompact 
to ultracompact size depending on the mass of the enclosed star or stars.
These small HII regions whose dynamics are
dominated by stellar gravitation and accretion are
different 
than compact and larger HII regions whose dynamics are dominated by
the thermal pressure of the ionized gas.
\end{abstract}


\section{Introduction} \label{introduction}

The hypothesis that massive stars form by accretion can be investigated
by  simple analytical calculations that describe the effect that the formation of
a massive star has on its own accretion flow. An earlier paper \citep{Keto2002b} studied 
the effect
of star formation in a spherically-symmetric, steady-state flow 
\citep{Bondi1952} and showed how
a molecular accretion flow
may pass through an ionization front and continue toward the star as an
ionized accretion flow. In this case, the HII region is the ionized inner zone 
of a continuous two-phase accretion flow. 
More generally, angular momentum in an accretion flow will result in a flattened 
geometry with different consequences for the structure of the ionized accretion flow and
HII region.  

This paper investigates  the
effects of the ionizing radiation of a massive star on
an accretion flow with angular momentum, modeled as
gas flow on ballistic trajectories around a star \citep{Ulrich1976}.
The model is described by a few parameters: 
the gas density in the flow, the angular momentum, the mass of the star, and the flux
of ionizing radiation. Alternatively, the model can described in terms of
three characteristic radii:
characteristic radii, $R_i$, the radius
of ionization equilibrium, $R_b$, the 
radius where the escape velocity equals the 
sound speed, and $R_d$, the radius of disk formation where the gravitational and
centrifugal forces balance,

Models with different ratios of these three radii produce HII regions with 
different morphologies and dynamics. 
In particular, increasing the ratio $R_i/R_b$ radius of ionization results
in a progression of HII regions from quasi-spherical, gravitationally-trapped
\cite{Keto2002b, Keto2003}
through bipolar to a final 
morphology similar to that around a
photo-evaporating disk as described by
\citet{Hollenbach1994}, \citet{Yorke1995}, \citet{YorkeWelz1996}, \citet{Lizano1996}, \citet{Johnstone1998}, and \citet{Lugo2004}. 
This progression suggests an evolutionary sequence for an HII region around
a star that is growing by accretion. As the star gains mass, the location
of the radius, $R_b$, will increase proportionally with the mass of the
star. The radius of ionization equilibrium, $R_i$, will also increase but
as a higher power of the mass because the flux of ionizing radiation
depends on the temperature of the star. Thus this simple model of
ionization within an accretion flow provides an understanding of the
relationship between the several different observed HII region
morphologies on the hypercompact and ultracompact scales.

\section{Relevant results from the spherical model: the 3 
evolutionary stages of an HII region}
\label{spherical}
Stellar structure models suggest that massive stars that form by 
contraction have a shorter
pre-main sequence phase (PMS) than low mass stars, and that massive stars
that form by accretion have no PMS phase at all
because massive stars begin hydrogen burning while still accreting 
\citep{PallaStahler1993, BeechMitalas1994, Chieffi1995,
BernasconiMaeder1996, BehrendMaeder2001, 
NorbergMaeder2000, Keto2003}. These calculations suggest
that a massive star that is forming by accretion will reach the main sequence with a mass
below that which would produce sufficient ionizing radiation to maintain an HII region
around the star. The non-existent HII region at this stage may be described as {\it quenched}
\citep{Walmsley1995}. Because the 
production of ionizing photons increases as a higher power of the stellar mass than does
the accretion rate, eventually the flux of
ionizing radiation, $J_\ast$, from a star that is gaining mass by accretion, 
will exceed  the flux of neutral gas  onto the star \citep{Keto2003},
\be
J_\ast > 4\pi r_\ast^2n_Hv
\e{quenched}
\ee
where $r_\ast$ is the stellar radius, $n_H$ is the the number density of neutral gas,
and $v$ the inflow velocity.

If the density of the accretion flow decreases with distance no faster than $r^{-3/2}$, then the HII
region that develops in the flow will be bounded at a distance, $R_i$
\citep{FrancoTTBodenheimer1990},
defined by the balance of the ionizations and recombinations \citep[][eqn. 5.20]{Spitzer1978},
\be
%\int^{R_i}_{r_0} {{dS(r)}\over{dr}} = S(R_i) - 
J_\ast - \int^{R_i}_{r_\ast} 4\pi r^2 n_e^2(r)\alpha^2 dr=0
%{{J_\ast}\over{4\pi R_{i}^2 }} - \int^{R_{i}}_{r_\ast} n_e^2 \alpha dr  = 0
\e{stromgren}
\ee
This equation incorporates the "on-the-spot" approximation
with $\alpha$ the recombination rate to all Rydberg levels of H above the first level, $n=1$.

The small HII region that  develops just as the quenched phase (equation \ref{eq:quenched}) 
ends 
cannot expand hydrodynamically because the outward
pressure of the ionized  gas is less than the inward gravitational force of the star. 
Instead, the ionized gas forms part of a continuous accretion flow with an outer molecular phase
and an inner ionized phase separated by a static ionization front within the flow.
The HII region, which at this stage is the inner part of the accretion flow, may be described as 
{\it gravitationally trapped} \citep{Keto2002b,Keto2003}. 
As the star continues to gain mass and its flux of 
ionizing radiation increases, the boundary of the HII
region, whose location is determined 
by ionization equilibrium rather than pressure balance, will be found at greater
radii. Once a radius is reached where the escape velocity from the star is below the
sound speed of the ionized gas, the HII region will begin to expand hydrodynamically.
This radius is approximately equal to the Bondi-Parker trans-sonic
radius where the velocity of
the incoming accretion flow first reaches the sound speed of the ionized gas,
\be
R_b = GM_\ast/2c^2
\e{bp}
\ee
where $M_\ast$ is the mass of the star and $c$ is the sound speed. 
The evolution of the HII region then transitions to a phase 
of {\it pressure-driven expansion} in which the gravitational attraction of the star
is negligible over most of the HII region. These pressure dominated HII regions
are adequately  described by models with no gravitational 
force \citep{Spitzer1978, DysonWilliams1980, Shu1992}. The model for the development
of an HII region within a spherical accretion flow thus suggests three stages: 1) quenched or
non-existent, 2) gravitationally trapped, and 3) pressure driven-expansion. The divisions
between the stages occur when the radius of ionization equilibrium, $R_i$, 
(equation \ref{eq:stromgren}).

\section{A non-spherical accretion flow} \label{nonspherical}

One simple model of accretion that includes rotation equates the streamlines of an accretion flow
with ballistic trajectories around a point mass 
\citep{Ulrich1976, CassenMoosman1981, Chevalier1983, Terebey1984}
The flow is determined by the gravitational attraction of the point source subject to conservation of angular momentum and mass.  In particular,  the model ignores 
the pressure and self-gravity of the gas. The initial distribution of angular momentum is
that of solid body rotation on an arbitrary radius or
conceptual spherical surface some distance from the
star. Thus on each trajectory, the specific angular momentum, $\Gamma \propto \sin{\theta_0}$,
where $\theta_0$ is the initial polar angle. The gas density is described by the conservation
of mass along streamlines. In this model,
an accretion disk develops as the gas density increases along inwardly converging streamlines
flattened by the angular momentum of the flow.
A limitation of the model is that the mid-plane density is not
defined  whereas a more complete model would describe the density across the mid-plane as
a pressure supported disk \citep{SS73,LBP1974,Pringle1981,Hartmann1998,Whitney2003}. 
This deficiency may be ignored if, as in the examples that follow, the scale of the
HII region developed within this accretion flow is larger than the scale height 
of the disk (see \S \ref{examples}).
The equations for the velocity and density at a position $(r,\theta)$ are \citep{Ulrich1976},
\be
v_r = -\bigg( {{GM}\over{r}} \bigg)^{1/2} \bigg( 1+ {{\cos{\theta}}\over{\cos{\theta_0}}} \bigg)^{1/2}
\e{vr}
\ee
\be
v_\theta = \bigg( {{GM}\over{r}} \bigg)^{1/2} (\cos{\theta_0} - \cos{\theta} )
	\bigg( 1+ {{\cos{\theta_0} + \cos{\theta_0}}\over{\cos{\theta_0} \sin{\theta} }} \bigg)^{1/2}
\e{vtheta}
\ee
\be
v_\phi = \bigg( {{GM}\over{r}} \bigg)^{1/2} 
	{\sin{\theta_0}\over{\sin{\theta}}}
	\bigg( 1+ {{\cos{\theta}}\over{\cos{\theta_0}}} \bigg)^{1/2}
\e{vphi}
\ee
where,
\be
r = { R_d\cos{\theta_0 }  \sin{\theta_0}^2 \over { \cos{\theta_0} - \cos{\theta}   }}
\e{radius}
\ee
This last equation includes the parameter, $R_d$, which is the radius at which the
gravitational force equals the centrifugal force in the mid-plane, 
\be
\Gamma^2/r^3_d = GM/r^2_d
\e{centrifugal}
\ee
This radius, approximately where the accretion flow 
transitions from a
quasi-spherical inflow to a rotationally supported disk, 
may also be written in a form
analogous to the Bondi-Parker transonic radius as,
\begin{equation}\label{eq:rdbp}
R_d = GM/v_k^2 
\end{equation}
where $v_k$, the orbital velocity at $R_d$, replaces the
sound speed in equation \ref{eq:bp}. 
The gas density from mass conservation is \citep{Mendoza2004},
\begin{equation}\label{eq:density}
n = n_0 r^{-3/2}\bigg(1+{{\cos{\theta}}\over{\cos{\theta_0}}}\bigg)^{-1/2}
(1 + r^{-1}(3\cos^2{\theta_0} - 1))^{-1}
\end{equation}
where the number density, $n_0$, is defined through the mass density, $\mu n_0$,
by the mass accretion rate,
\begin{equation}\label{eq:accretion}
\dot{M} = \mu n_0 4\pi R_d^2 v_k
\end{equation}

\section{The development of an HII region in the non-spherical accretion flow} \label{developmentHII}

An HII region will develop at the center of  the flow defined by 
equations \ref{eq:vr} through  \ref{eq:accretion} once the ionizing flux of the
accreting star  satisfies the condition in equation \ref{eq:quenched}.  
However, in a non-spherical flow,
the gas density  is not radially
symmetric, and therefore the HII boundary, $R_i$, is
at larger distances in directions where the
gas density is lower. In the axially symmetric model above, the boundary is therefore
a function of the polar angle, $\theta$.
The spherical model of \S\ref{spherical} suggests that if $R_i > R_b$ the HII region expands
hydrodynamically. 
Thus while somewhat idealized, in the non-spherical model
the HII region may be expanding over a range of angle $\theta$ about the polar axis  
while the accretion flow
continues to flow into and through the HII in a different range of angle $\theta$ around the plane of the disk.

This simple model of the ionization of an accretion flow does not explicitly include the effect
of a stellar wind on the HII region. We know that unembedded
early-type stars produce
powerful winds that at their terminal velocities have mechanical energies, $\rho v_t^2$,
that exceed the mechanical energies of star-forming accretion flows
\citep{Lamers1999}.
However, because the
winds accelerate off the surface of the star the wind energy is negligible near the 
star and, in the idealized spherical
model, the wind may be suppressed
by the accretion flow. Once inflow ends because the gas within the HII region begins 
expanding ($R_i > R_b$),
the wind may be expected to break  out
and dominate the hydrodynamics. Thus in the
non-spherical model, the following approximation is adopted. At a polar angle, $\theta$,
where $R_i (\theta) > R_b$, the inflow is replaced by the simplest model of a stellar
wind (Parker 1958). This wind solution is the analogue of the Bondi accretion flow and
described by the same equation,
\be
{{d}\over{dr}}\bigg({{v^2}\over{2}}\bigg) +{{dP}\over{dr}} + {{GM}\over{r^2}} = 0
\e{wind}
\ee
but with the boundary conditions that the flow is subsonic at the stellar surface and
supersonic beyond the transonic radius. In the case of non-spherical flows, the location 
of the transonic radius depends on the geometry of the flow, specifically on the divergence
of the streamlines about the star \citep{KoppHolzer1976}. However, if the divergence is approximately 
spherical,  the transonic point will be approximately located by equation \ref{eq:bp}.
For simplicity, this approximation is adopted. The resulting model of a wind in the
polar directions and accretion at lower latitudes is inconsistent in that the wind solution
is driven by hydrodynamic pressure, but 
pressure is neglected
in the accretion flow. 
Nonetheless, the examples in \S\ref{examples} will show that this approximation is useful 
on a conceptual level and produces models that describe the morphology of observed
HII regions. Different models that describe the interaction of a wind and accretion
flow include \citet{WilkinStahler1998}, \citet{Mendoza2004}, and \citet{Cunningham2005}.

\section{The 3 R's of hypercompact HII regions} \label{3Rs}
The equations in sections  \S\ref{spherical} and \S\ref{developmentHII} fully
describe the model of an HII region in
a non-spherical accretion flow. The morphology of the HII region depends on
a few physical quantities that are specified as parameters of the model: 
the gas density and angular momentum
in the accretion flow, and the mass and flux of ionizing radiation of the star. Because these
physical quantities have different units, it is helpful to organize them into 
parameters of the same physical dimension as 3 characteristic radii.  
The radius of disk formation, $R_d$ (equation \ref{eq:rdbp}), 
along with the Bondi-Parker radius, $R_b$
(equation \ref{eq:bp}), and
the radius of ionization equilibrium, $R_i$ (equation \ref{eq:stromgren}), 
may be called the 3 R's of HII 
regions in that they describe the basic structure of the accretion flow and HII region.  

We may also combine these 3 characteristic radii into non-dimensional ratios, for example
$R_i/R_d$, $R_i/R_b$, $R_d/R_b$ to see that the models are scale-free, and that
it is the relative magnitudes of the 3 radii that are important in describing the 
structure of the accretion flows and HII regions. Thus a model
appropriate for a common accretion flow onto several
O stars such as in G10.6-0.4  with a total stellar mass of a few 
hundred M$_\odot$ \citep{KetoWood2006}
may have the same structure as a model for a single B star 
such as IRAS20126 \citep{Cesaroni1997, Cesaroni1999} provided that the 
three ratios are the same.

Because the model structures depend on only 3 characteristic radii or 3 ratios,
then only a few model examples are required to illustrate the possible
variations as well as suggest an evolutionary sequence for HII regions.
So that these models may be more easily related to the densities and stellar types
familiar from observations of high mass star forming regions, the examples in the
next section
are presented in units appropriate for early B stars and late O stars since these are most
commonly observed massive stars. These models are applicable, if scaled up, to accretion flows
as large as those onto groups of early O stars such as G10.6-0.4 \citep{KetoWood2006}.

\section{Examples and evolution} \label{examples}
\subsection{Ionization}\label{ionization}

The first  example
illustrates the effect of increasing the ionizing flux within the same accretion flow. 
This comparison illustrates the differences as $R_i/R_b$ is 
greater than, approximately equal to, or less than one. 
(In this example for the purpose of this comparison,
we set $R_d \approx R_b$.)  
As the ionizing radiation from the star increases, and $R_i$
increases relative to $R_b$, the HII region
transitions through three phases from nearly spherical inflow of the ionized gas through
bipolar to nearly spherical outflow. 

The initial model is composed of a B1 star with a mass of 18 M$_\odot$ and
a flux of ionizing radiation of $10^{45}$ photons s$^{-1}$ \citep{Vacca1996} within an 
accretion flow with 
$R_d = 42$ AU ($\Gamma_0 = 0.04$ kms$^{-1}$ pc), $R_b = 43$ AU,  and a gas density, $n_0$ of $10^7$ cm$^{-3}$. 
The mass accretion rate (equation \ref{eq:accretion}) is $6\times 10^{-6}$ M$_\odot$ yr$^{-1}$.
Stellar structure calculations suggest that 
this accretion rate is near the minimum required for the formation of stars of this mass
\citep{Keto2003, KetoWood2006}. Higher rates might be required to form stars of higher mass.

Figure \ref{fig:ionizingflux} ({\it top}) shows that
at this flux
level and density,  although the HII region has a 
slightly bipolar shape owing to the flattening of the flow, the boundary of the HII region is
entirely with the Bondi-Parker critical radius
($R_i < R_b$ at all polar angles). Thus the HII region is trapped within
the gravitational field of the star, and accretion may proceed through the HII region at all
angles. 

Increasing the stellar mass to 20 M$_\odot$ and the flux of ionizing radiation to
$7\times 10^{45}$ photons s$^{-1}$, corresponding to an B0.5 star, increases the extent
of the ionization sufficiently so that $R_i \geq R_b$ in a narrow range of angle about
the bipolar axis. Within this angular range, we assume that the HII region begins to expand
and the structure and dynamics are described by an outflow or wind driven by the
thermal pressure of the ionized gas. Accretion through the HII region
continues at lower latitudes (figure \ref{fig:ionizingflux} {\it middle}).

Increasing the stellar mass to 22 M$_\odot$ and the flux of ionizing radiation to
$10^{47}$ photons s$^{-1}$, corresponding to an O9 star, results in a model with $R_i > R_b$,
and thus outflow,
everywhere except at a narrow range of angle just around the disk 
(figure \ref{fig:ionizingflux} {\it bottom}). 
This third-stage structure is
similar to the photo-evaporating disk model
\citep{Hollenbach1994, Yorke1995, YorkeWelz1996, Lizano1996, Johnstone1998, Lugo2004}
except that here the outflow is radially 
from the center of the HII region rather than vertically
off a disk. A more complete model would include the evaporation off the disk
in the outflow.

In this scenario, 
the bipolar phase is relatively brief in the evolution between
the gravitationally trapped and outflow phases. In this
example,
the bipolar phase corresponds to a range 
of stellar mass between 20 and 22 M$_\odot$.

% .run init

% density = 1. e7
% arrow scale 0.3
% rotation 0.2 e5 at 0.0006 pc = 123 AU

% init, AngMom='low',fluxlevel='low'
% flux level = 1. e45
% mass = 18 Msun
% Rd = 42 AU
% keplerian velocity 19 kms
% Bondi radius 43
% plot scale 2.92 Rd

% init, AngMom='low',fluxlevel='high'
% flux level = 1.e47
% mass = 22 Msun
% Rd = 34 AU
% keplerian velocity 23 kms
% Bondi radius 59
% plot scale 3.57 Rd

\subsection{Angular momentum}\label{angularmomentum}
The second example shows the effect of different values of the 
angular momentum on the structure
of the accretion flow. This example illustrates the difference as
$R_b/R_d$ is greater or less than one.
Flows that have relatively low angular momentum will produce
what may be described as a fat accretion torus around a growing star while those
with relatively high angular momentum will produce a thin disk. Because the density
of the flow increases inward following mass conservation regardless of the angular
momentum, the difference between the two cases relates to where in the flow the 
gas density becomes observationally significant with respect to the flattening of the
flow. The flow becomes observable as the gas
density approaches the critical density for collisional de-excitation,
%\be
$n_c = A_{ij}/C_{ij}$,
%\e{critical density}
%\ee
where $A_{ij}$ is the Einstein A coefficient (sec$^{-1}$)
and $C_{ij}$ is the collision rate (cm$^{-3}$ sec$^{-1}  $)
for the particular
molecular tracer employed in an observation. 
In a model with low angular momentum,
the gas will reach the critical density before flattening into a disk, and the molecular
accretion flow will be detected as a quasi-spherical rotating flow or fat torus. In the
high angular momentum case, the flow will form a disk before the gas density in the
surrounding flow exceeds the critical density. In this case the observations will 
tend to see the flattened disk.

Figure \ref{fig:figure2} shows two variations of the same model. 
Both models include a star of 20 M$_\odot$ and flux of ionizing radiation of 
$3\times 10^{46}$ photons s$^{-1}$ corresponding to type B0 -- O9.5.
At this mass, $R_b = 54$ AU. In one model,
the angular momentum is 
$\Gamma_0 = 0.04$ kms$^{-1}$ pc,  while in the second model
the rotation rate is 2 times higher. The
values of $R_d$ are therefore 38 AU and 152 AU respectively. 
Because of the
greater inward increase in the gas density in the flattened, high-angular-momentum flow, the
density in this flow is set  lower, $n_0 = 1\times 10^6$ cm$^{-3}$,  rather
than $n_0 = 3\times 10^7$ cm$^{-3}$, so that similar densities are maintained at the center
of the flow, and therefore, similar opening angles for the bipolar outflow. 
The mass accretion rates for each
model are $6.6\times 10^{-6}$ and $1.8\times 10^{-5}$ M$_\odot$ yr$^{-1}$.
The two
density distributions shown in figure 2 suggest that
accretion disks and "fat toroids" may be
different expressions of the same model accretion flow with different values of
the angular momentum or equivalently different values of  $R_d$ with respect to $R_b$.

% B0.5  M=18.4  Q=45.69
% B0     M=19.5  Q=46.23
% O9.5 M=20.8  Q=46.77
% O9     M=22     Q=47.25

%.run init_large_scale
% Bondi radius 54 AU
% stellar mass = 20
% sound speed of Parker wind is 5000 K
% flux level 3. e46

%init,AngMom='low',fluxlevel='med'
% density 2.85 e7
% arrow scale = 2.5
% arrow scale outflow = 0.6
% rotation = 0.2 e5 at 0.003 pc 
% Rd 38 AU
% Keplerian velocity 21 kms
% plot scale 16.2  Rd

%init,AngMom='high',fluxlevel='med'
% rotation = 0.4 e5 at 0.003 pc
% arrow scale = 0.3
% arrow scale outflow = 0.3 
% density 1.00 e7
% Rd 152 AU
% Keplerian velocity 10.8 kms
% plot scale 4.05 Rd

\subsection{Limitations of the model}

The model of accretion on ballistic trajectories
in section \S \ref{nonspherical} does not include pressure forces
and therefore does not describe a pressure supported disk at the
mid-plane. The density in such a disk could 
potentially affect the morphology of an HII region depending on
the structure of the disk.
Little is known either observationally or theoretically
about disks around massive stars. 
If these disks are similar
to those around low mass stars, then we can calculate
how a scaled-up, low-mass disk would affect the
HII region morphologies in the examples above.
To do this we add to the model accretion flows in the examples above the density of 
a disk described as \cite[][equation 3]{Whitney2003},
\begin{equation}
\rho(r) = \rho_0 
\bigg( 1 - \sqrt{ {{R_*}\over{r}} }
\bigg)\bigg({{R_*}\over{r}} \bigg)^\alpha
\exp \bigg[ -{{1}\over{2}}\bigg( {{z}\over{H(r)}}   \bigg)^2 \bigg]
\end{equation}\label{eq:disk}
where $r$ is the radius in the disk mid-plane, $z$ is the height
above the plane,
$\alpha = 2.25$,
and the scale height, $H(r) = H_0(r/R_*)^\beta$ with $H_0 =0.1 R_*$ (AU)
and $\beta = 1.25$,
The density, $\rho_0$ 
is defined by the assumption that the total mass of the disk 
within $R_d$ is $0.1M_*$.
These additional calculations (not shown)
demonstrate that the inclusion of such a disk 
does not affect
the morphologies of the HII regions in these examples. 
This is because the scale height
of the pressure-supported disk at the radii where the boundary of the
HII region meets the mid-plane is small compared to the thickness of
the "disk" created by the flattening of the accretion flow. Because 
this pressure-supported disk is an arbitrary addition to the model
that does not affect the results sought in this investigation, this
disk is not considered further.

\section{The importance of stellar gravity and accretion on Hypercompacts
and small Ultracompacts}

The morphologies produced by models of HII regions that develop within
accretion flows are potentially quite varied and match many of the observed
morphologies, particularly those observed at the smallest scales. In particular, 
the bipolar HII regions are predominantly a small scale phenomenon. For example, 
there is no classification for bipolar morphology in the \citet{WoodChurchwell1989} and
\citet{Kurtz1994} survey of ultracompact HII regions whereas
\citet{Depree2005} find it necessary to introduce this class to describe the
morphologies of hypercompact HII regions. 

The size scale, $\sim  0.01$ pc (2000 AU), of a large 
hypercompact or small ultracompact HII region \citep{Kurtz2000}, 
is about 20 times the radius, $R_b$ around
a single massive star. What then is the relevance of the stellar 
gravity for most observed HC and UC HII regions? 

First, O stars do not appear to form alone, but in small groups
or clusters of early type stars 
that may also contain some number of lower mass stars. The radius, $R_b$,
scales with the attracting mass. Groups or small clusters of stars with 
a few hundred M$_\odot$ contained within a common HII region are suggested by
observations of a few of the brightest HII regions that show
molecular or ionized accretion on scales of $10^3$ AU 
\citep{ZhangHo1997, Young1998, Keto2002a, Sollins2005a, KetoWood2006}.  
Thus the gravity of groups or small clusters
may be significant even on the size scales of
ultracompact HII regions. 

Second, the HII region need not be strictly smaller than $R_b$ for 
the stellar gravitational attraction to affect its structure. 
For example, if the HII region is in effect an expanding wind, as in the model shown in figure 
\ref{fig:ionizingflux} ({\it bottom}), then because the wind solution is quasi-hydrostatic  
\citep[][equation 4]{Keto2003}
inside of $R_b$,  the HII region will be quite dense within this radius even though
the boundary, $R_i > R_b$. 
Outside of $R_b$, mass conservation within the converging accretion flow and within
the diverging outflow generally requires that the flows have density gradients.
Thus even beyond $R_b$, in the region dominated by pressure forces, the density is
not uniform.
High frequency
observations would show a
bright core 
whereas
low frequency observations of the same HII region would show 
surrounding low-level extended emission.

\section{Jets, outflows and disks}

The outflows described here are simple models of stellar winds 
driven by the 
pressure of the ionized gas (Parker isothermal wind \citep{Parker1958})
in the HII region and confined in a
specific way defined by the ratio of $R_i/R_b$ as a function of angle, 
but nonetheless essentially by the
geometry of the surrounding accretion flow. Thus these outflows require both 
significant ionizing flux from the star and a massive accretion flow. In this respect
they are quite different from the more familiar bipolar jets or outflows associated with 
young low-mass stars. These latter are not completely understood, but are
thought to be driven by the twisting of magnetic fields by a thin accretion disk.
Although this paper has not dealt with magnetically-driven outflows, these 
are not incompatible with the HII-driven outflows. 
The hypothesis of massive 
star formation presented here, following \citet{Keto2002b, Keto2003},
suggests that both occur. In this hypothesis, high mass stars grow by accretion from lower mass
stars. Since magnetically-driven outflows are
an inescapable part of low-mass star formation, the massive stars-to-be must reach
spectral type B with a magnetically-driven outflow. 

As the star continues to grow to earlier type B, and
the ionizing radiation creates an HII region,
what becomes of the magnetically driven outflow?  \citet{BS2005} suggest that
there is a of lack observations of collimated jet-like outflows associated with
stars earlier than type B1 compared with the number of observations of jet-like outflows
from later type B stars. At the moment we do not know whether the jets are 
destroyed by processes associated with the ionization or
survive the formation of the HII region \citep{TanMcKee2003} but are difficult to detect.

\section{Comparison with observations}

A model of accretion with a bipolar ionized outflow
has been compared to observations of both ionized gas and molecular gas in and
around the ultracompact HII region
G10.6-0.4  \citep{KetoWood2006}. 
The model for the accretion flow in G10.6-0.4 is based on the same model
of streamlines on ballistic trajectories presented here.
These observations are unique in
mapping the velocities in both the molecular and ionized gas and in following the
accretion flow from the molecular phase through the ionized phase. However, the source
G10.6-0.4 is not unique.
Molecular line observations of G28.20-0.05 \citep{Sollins2005b} and G24.78+0.08 \citep{Beltran2006}
have also been interpreted as accretion 
disks and outflows.  
Other observations that report disks or torii around massive stars, some with associated
outflows include \citep{Cesaroni1997, Hofner1999, Cesaroni1999,
ShepherdKurtz1999, Shepherd2001, Beltran2004, 
Beltran2005, Zhang1998, Zhang2002,
 Kumar2003, Chini2004, Beuther2004, Patel2005}

\section{Conclusions}

This paper provides a simple theoretical description of 
some of the effects of the ionization of a massive-star forming accretion flow. 
The model presented here is based on three simple models:  (1) accretion with
rotation as streamlines on ballistic trajectories \citep{Ulrich1976};
(2) an outflow as a pressure-driven 
isothermal wind \citep{Parker1958};
and (3) an ionization front gravitationally trapped within an accretion flow \citep{Keto2002b}.
The composite model applies to the evolutionary phase in the formation of massive stars 
of type B and earlier when the stars are both hot enough to ionize an HII region 
around the star and yet are still growing by accretion.
The model accretion flows and HII regions may be
described in terms of three characteristic radii which completely
determine the 
morphology and flow pattern. Variation of the relative magnitudes of the
parameters suggests how different morphologies may be related to
the same underlying model. The models further suggest 
an evolutionary sequence driven by the increasing ionization of 
a star growing by accretion. 
These hypercompact 
to ultracompact HII regions are
different in that their dynamics are dominated by gravitational force of their stars
whereas the larger HII regions are dominated by the thermal pressure of the
ionized gas.

\vbox{
{~}\hfill\break
\includegraphics[angle=90,width=4.0truein]{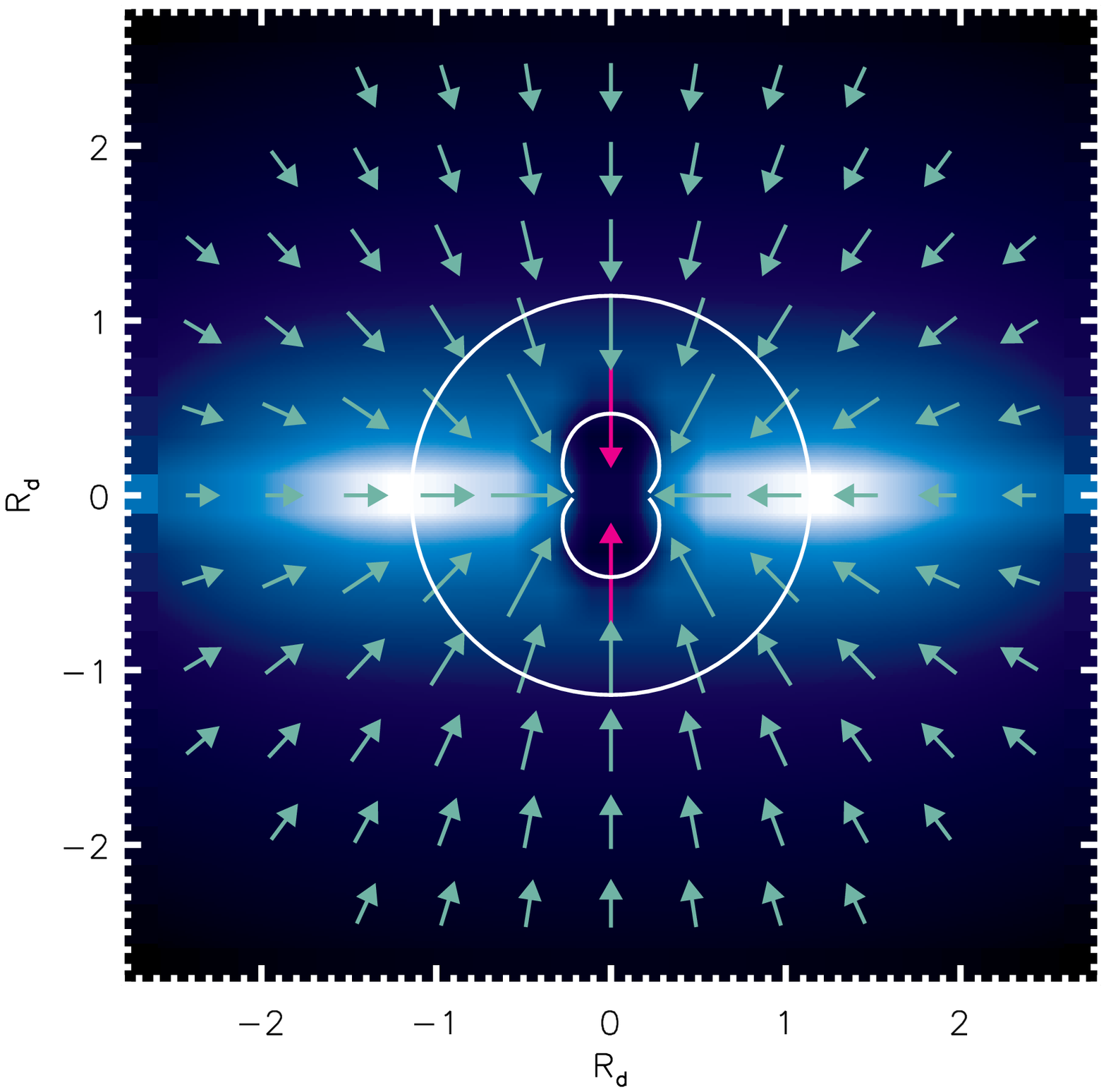}\hfill\break
\includegraphics[angle=90,width=4.0truein]{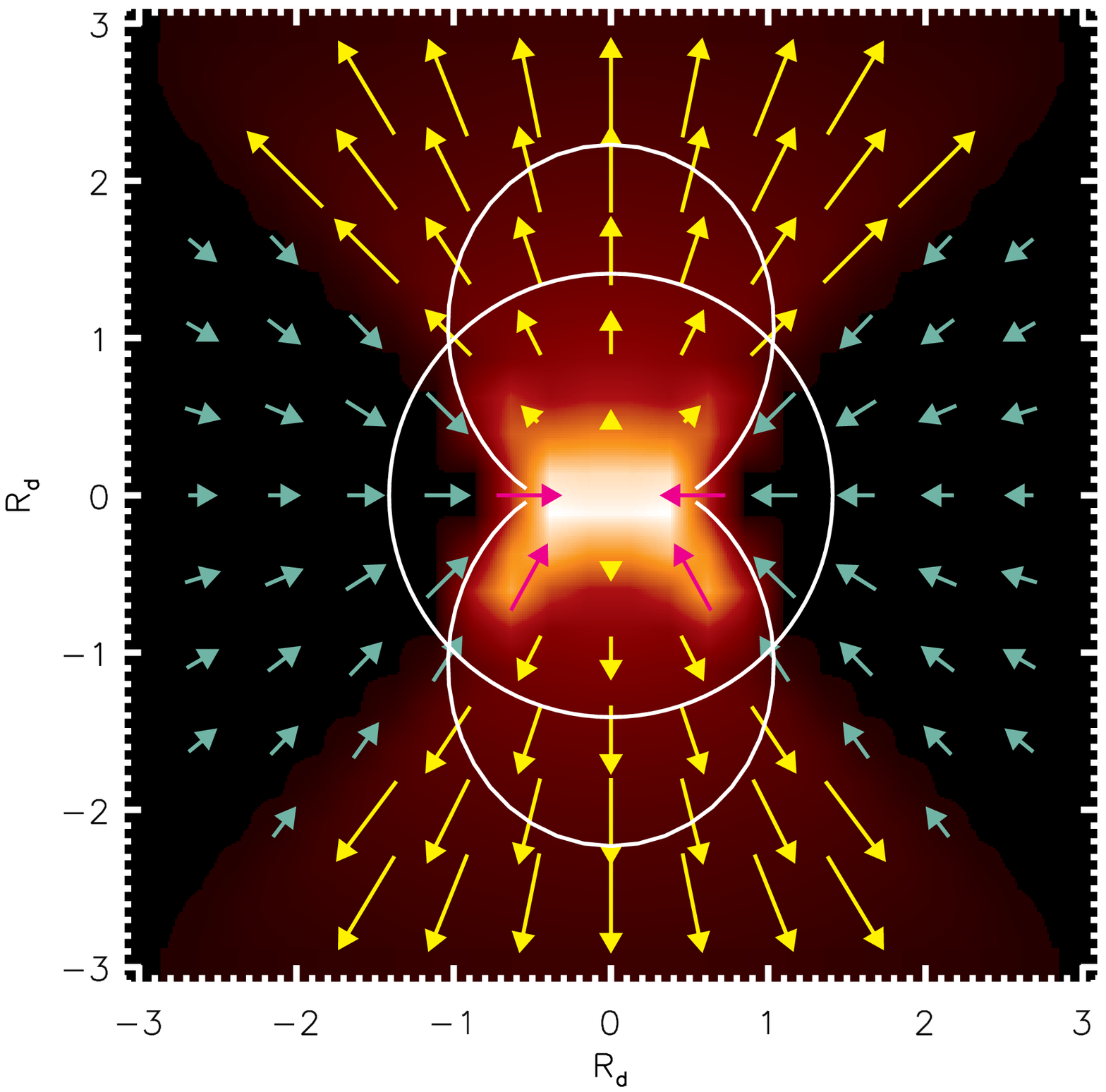}\hfill\break
\includegraphics[angle=90,width=4.0truein]{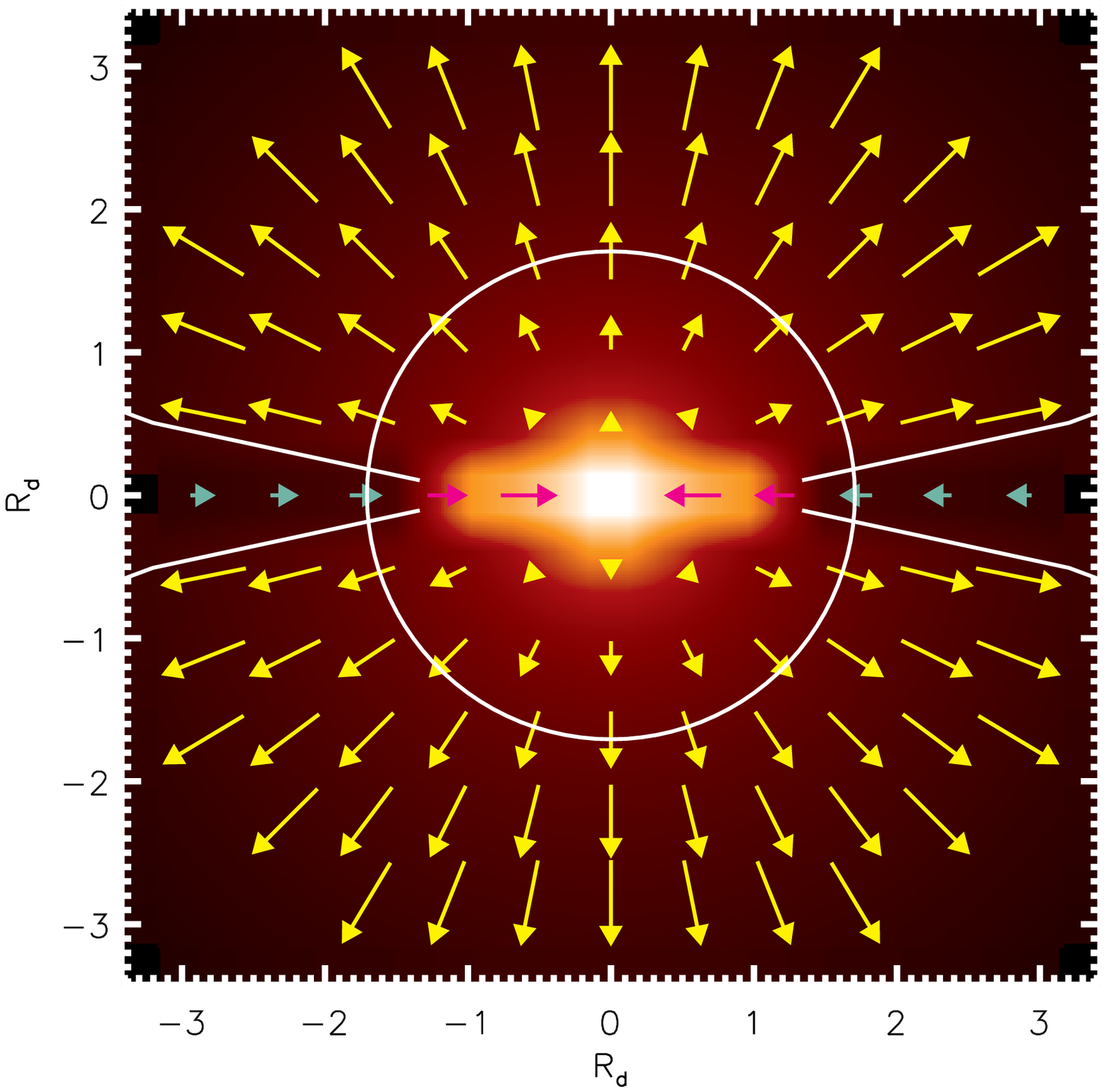}
}
\clearpage
\begin{figure}
\caption{ Model of a high angular momentum accretion flow subject to 3 levels
of ionizing radiation, low ({\it top}), medium ({\it medium}), 
and high ({\it bottom}) as defined in  \S\ref{examples}. 
The figures show the log of the
density of molecular gas in blue ({\it top}) and of the ionized gas in
red ({\it middle {\rm and} bottom}) in a slice in the XZ plane of
the flow. The color scales range from 0 to $1.6\times 10^7$ cm$^{-3}$ 
(molecular) ({\it top}),
from 0 to $1.2\times 10^7$ cm$^{-3}$ (ionized) ({\it middle}),
and 0 to $1.3\times 10^7$ cm$^{-3}$ (ionized) ({\it bottom}). 
The circle shows the location of the Bondi-Parker critical radius of the ionized
gas for spherical
flow. The arrows show the velocity
of the flow in the XZ plane. 
In the top figure, the longest arrow in the molecular flow represents 26.6 kms$^{-1}$
and the longest arrow in the ionized flow represents 21.5 kms$^{-1}$. 
In the middle figure, the longest arrow in the molecular flow represents 8.0 kms$^{-1}$
and the longest arrow in the ionized flow represents 28.2 kms$^{-1}$. 
In the bottom figure, the longest arrow in the molecular flow represents 5.4 kms$^{-1}$
and the longest arrow in the ionized flow represents 29.4 kms$^{-1}$. 
In the ionized outflow flow, the velocity is the sound speed at the critical radius.
The axes are labeled in units of $R_d$, 42 AU ({\it left}) and 34 AU  ({\it right}).}
\label{fig:ionizingflux}
\end{figure}
\clearpage

\includegraphics[angle=90,width=3.5truein]{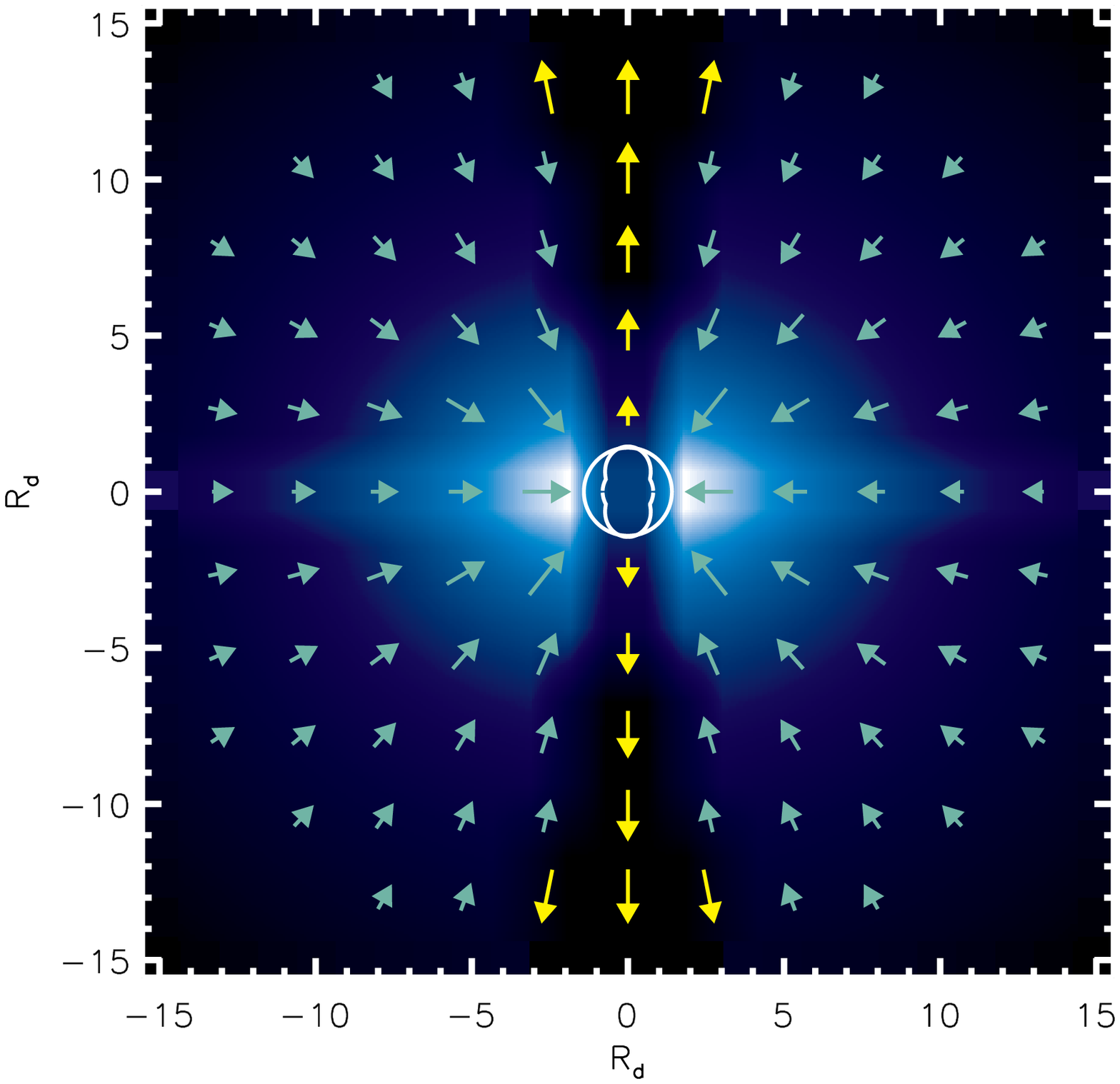}
\includegraphics[angle=90,width=3.5truein]{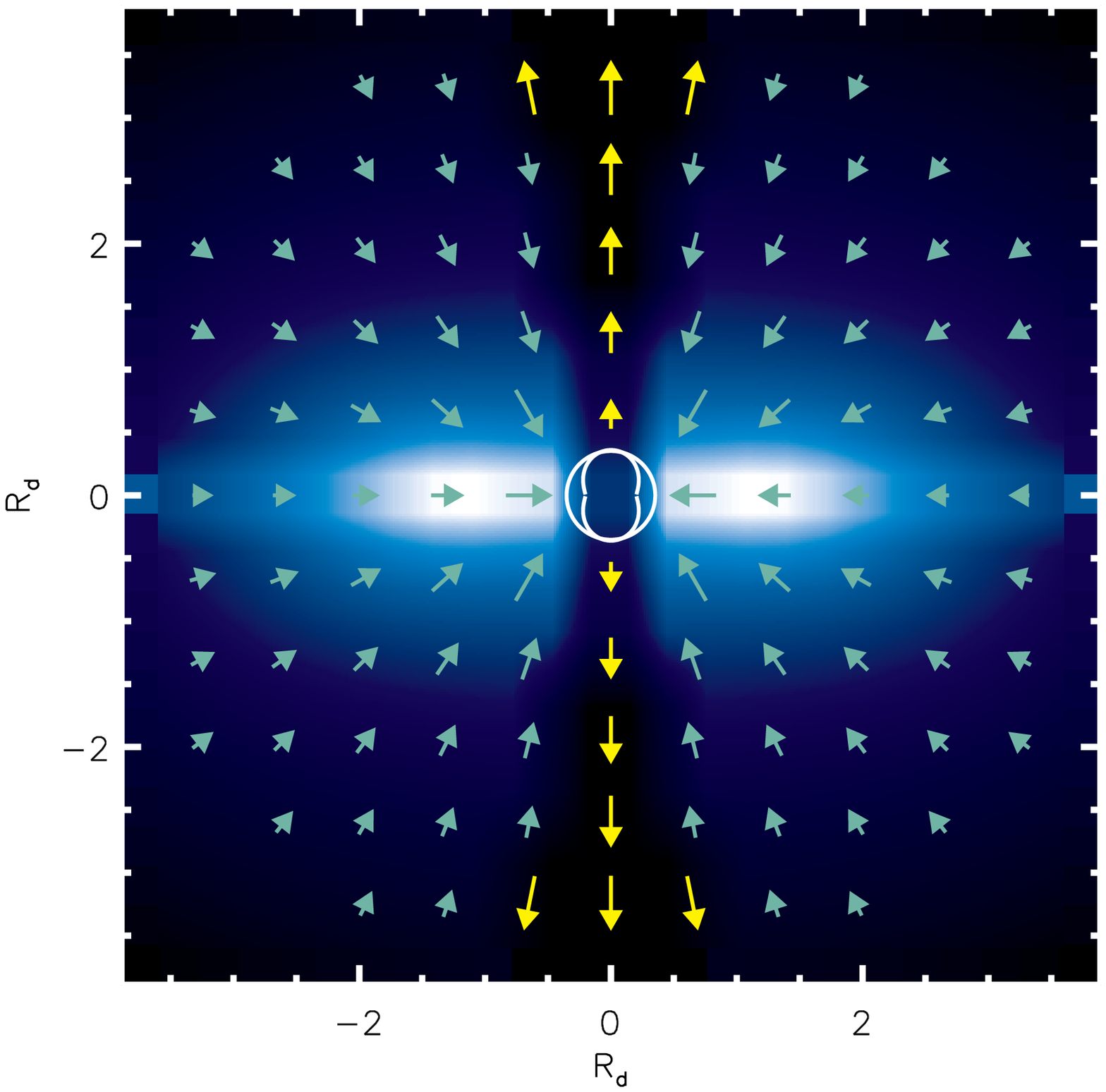}
%\vskip 1.0truein
\begin{figure}
\caption{Two accretion flows that differ in the relative amount of angular
momentum. The flow in the right column has 2 times the angular momentum of the
flow in the left column. The two figures show a slice in the XZ plane with 
the log of the density of the molecular gas in color and the velocity  as arrows. 
The density ranges from  $4.5 \times 10^3$ to 
$4.5\times 10^6$  cm$^{-3}$ (molecular) ({\it left}) and
from $4.5\times 10^3$ to $5.2\times 10^6$ cm$^{-3}$  ({\it right}).
The longest arrow in the  molecular flow (blue arrows) 
represents 1.8 kms$^{-1}$ ({\it left}) 
and   0.4 kms$^{-1}$ ({\it right}).
The yellow arrows show the velocity of the ionized outflow at half the scale. 
}
\label{fig:figure2}
\end{figure}
\clearpage

\end{document}